\documentclass[11pt,twoside]{article}

\usepackage{asp2006}
\usepackage{epsf}
\usepackage{psfig}
\usepackage{lscape}
\usepackage{graphicx}

\begin{document}
\title{The Formation of Hydrogen Deficient Stars through Common Envelope Evolution}   %%% Fill in title
\author{Steven Diehl, Chris Fryer}   %%% Fill in author names
\affil{Los Alamos National Laboratory, P.O. Box 1663, Los Alamos, NM-87545, USA}    %%% Fill in author affiliations
\author{Falk Herwig}   %%% Fill in author names
\affil{Astrophysics Group, School of Physical and Geographical Sciences,
Keele University, UK}    %%% Fill in author affiliations

\begin{abstract} %%% Abstract to run on from here.

We present preliminary results from Smooth Particle Hydrodynamics (SPH) simulations of common envelope evolution. We qualitatively compare the interaction between a $0.9M_{\sun}$ red giant with two different companion masses: a $0.05M_{\sun}$ brown dwarf and a $0.25M_{\sun}$ white dwarf companion. 

\end{abstract}

%%% MAIN BODY OF TEXT GOES HERE. CONSULT "INSTRUCTIONS FOR AUTHORS USING
%%% LATEX2E MARKUP", SECTIONS 2.3-2.6 FOR HELP WITH EQUATIONS, FIGURES,
%%% AND TABLES.
%\section{}   %%% Top level section head (remove "%" symbol)
%\subsection{}   %%% Second level section head (remove "%" symbol)
%\subsubsection{}   %%% Lowest level section head (remove "%" symbol)
%\section*{}    %%% Unnumbered top level section head (remove "%" symbol)
%\subsection*{}   %%% Unnumbered second level section head (remove "%" symbol)

\section{Introduction}

Most stars (60$\%$) are part of binary or multiple systems. If the binary separation is small enough, the more massive star will eventually engulf the companion as it expands during its giant phase, and form a so-called common envelope (CE) system. Friction then transfers orbital energy and angular momentum into the envelope, which may eventually lead to ejection of the entire envelope. Most tight binary systems are believed to have been in a CE phase at one point in their life: high- and low-mass X-ray binaries, double degenerate white dwarf and neutron star binaries, cataclysmic variables and supersoft X-ray sources. Many stellar types -- such as hydrogen deficient stars, early R-stars, or Wolf-Rayet type central stars of planetary nebulae for example -- have also been proposed to be the result of a CE phase, as isolated stellar evolution models are not able to fully explain their properties \citep[e.g.][]{DeMarcoWolfRayet, IzzardRstars}. New studies have even proposed that the majority of planetary nebulae may be the result of a CE interaction \citep{DeMarcoPNbinaries}.
 
The concept of CE evolution has been around since the pioneering work of \citet{PaczynskiCE}. Earlier simulation efforts \citep{RasioCE, SandquistCE} were restricted to high mass companions due to the low numerical resolution achievable at that time. Only a handful low-resolution simulations are available as of now, covering only a tiny fraction of the vast parameter space available. Only very recently, \citet{RickerCE_AMR2} have restarted simulating these dynamic events at high resolution with the adaptive mesh refinement code {\it FLASH}, so far concentrating again on interactions with relatively high-mass companions. Theoretical considerations expect lower-mass companions (such as brown dwarfs, or even Jupiter-mass planets) to penetrate much deeper into the giant's envelope or even merge into the core, requiring a much higher numerical resolution at the center, \citet[see e.g.][]{DeMarcoWolfRayet2}. Yet, these binaries should be much more common and may easily prove powerful enough to shed the rather losely bound envelope of a red giant (RG) or asymptotic giant branch (AGB) star. Here we will present a qualitative comparison between two CE evolution simulations involving a $0.9M_{\sun}$ red giant interacting with a low ($0.05M_{\sun}$ brown dwarf, BD) and medium mass ($0.25M_{\sun}$ white dwarf, WD) companion. The BD-RG system was chosen to be the progenitor system of a recently discovered tight binary system between a $0.05M_{\sun}$ BD and a $0.39M_{\sun}$ WD \citep{MaxtedRGBD}. Our RG was chosen such that its degenerate core matches this WD mass. We will use this system as a validation case for our code, and thus expect the BD to spiral in as far as the observed $0.65R_{\sun}$ separation.

\begin{figure}[t]
\hfil
\includegraphics[width=0.25\textwidth]{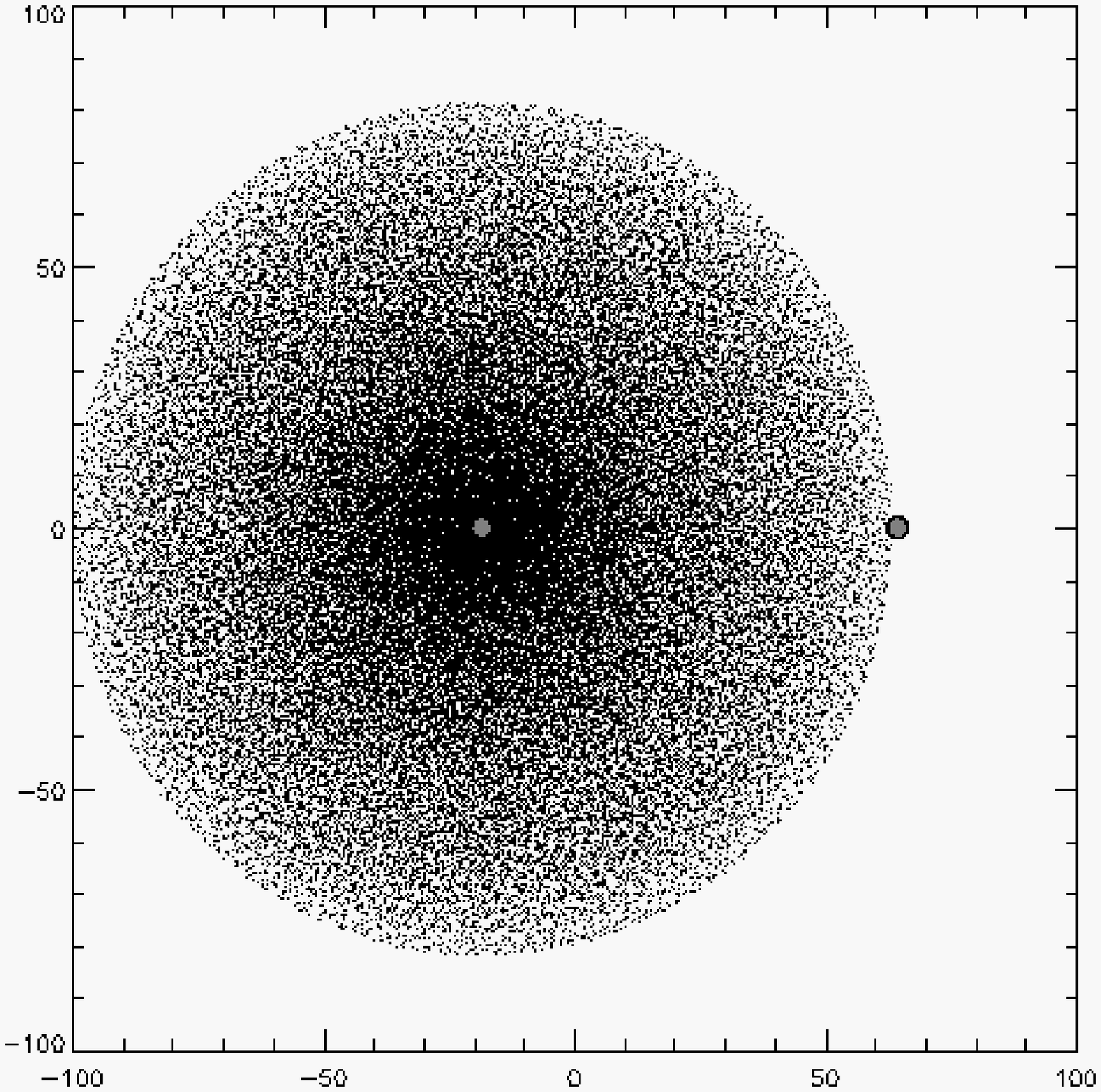}
\includegraphics[width=0.25\textwidth]{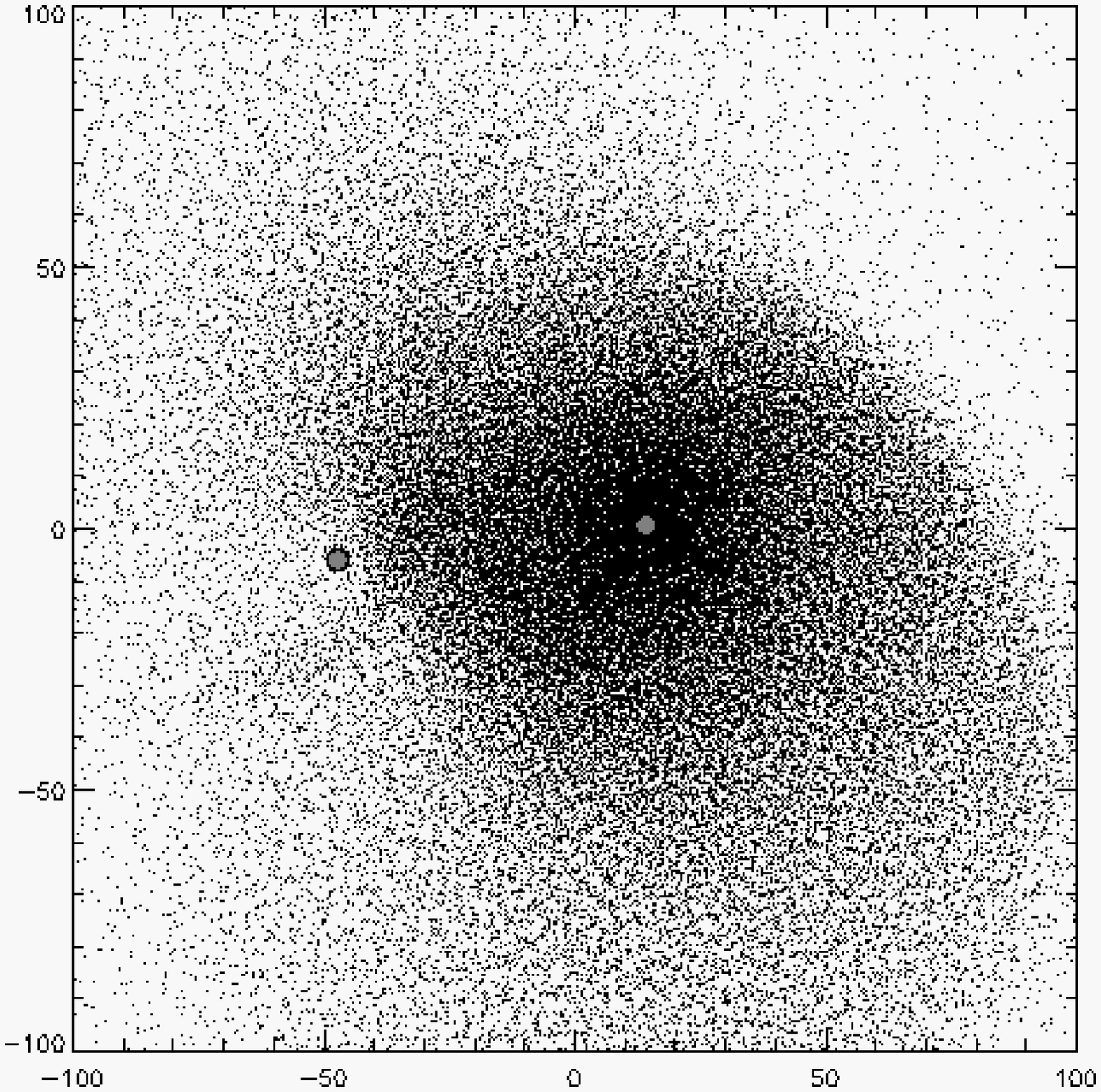}
\includegraphics[width=0.25\textwidth]{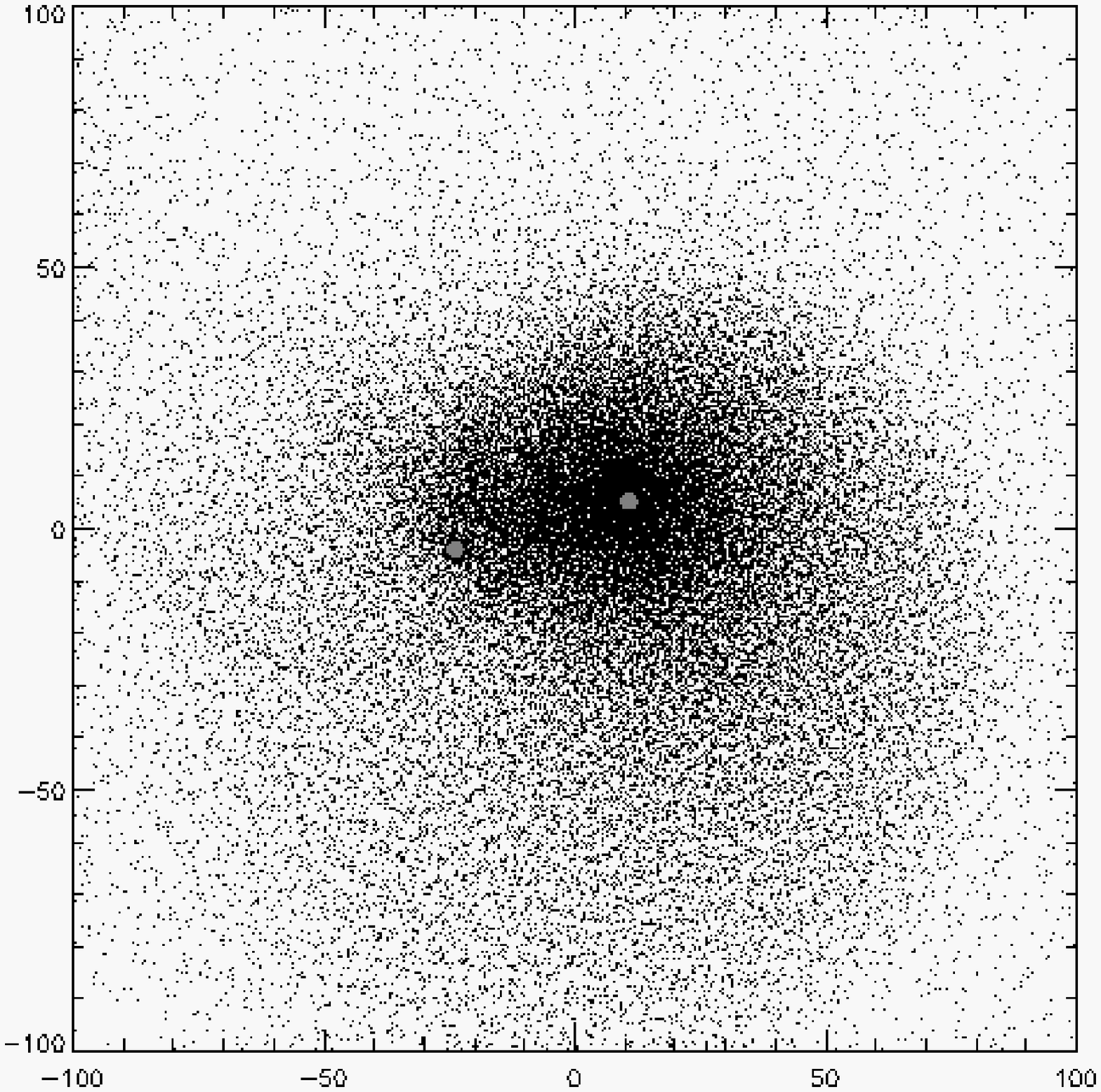}
\hfil \\
\null\hfil 
\includegraphics[width=0.25\textwidth]{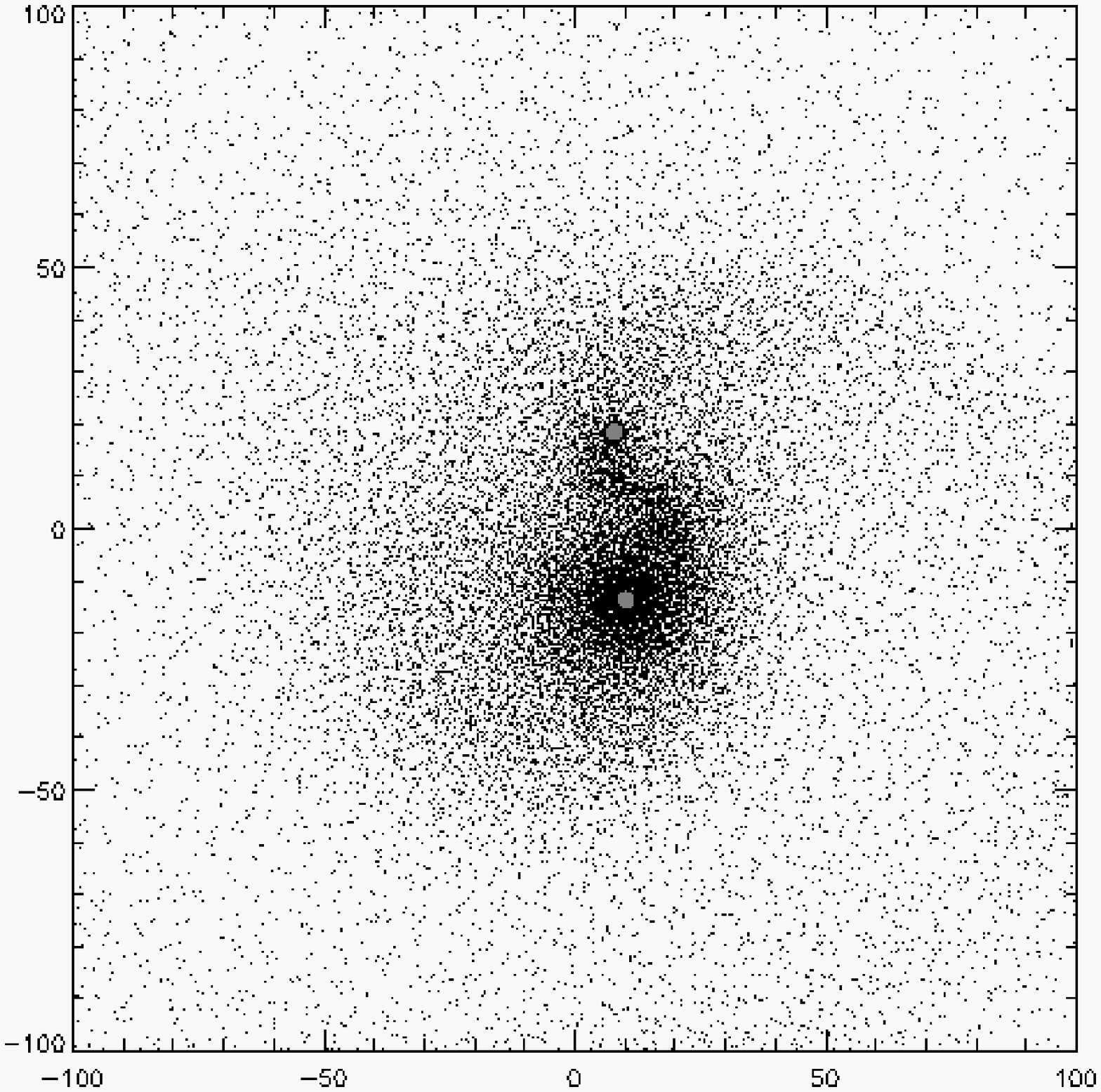}
\includegraphics[width=0.25\textwidth]{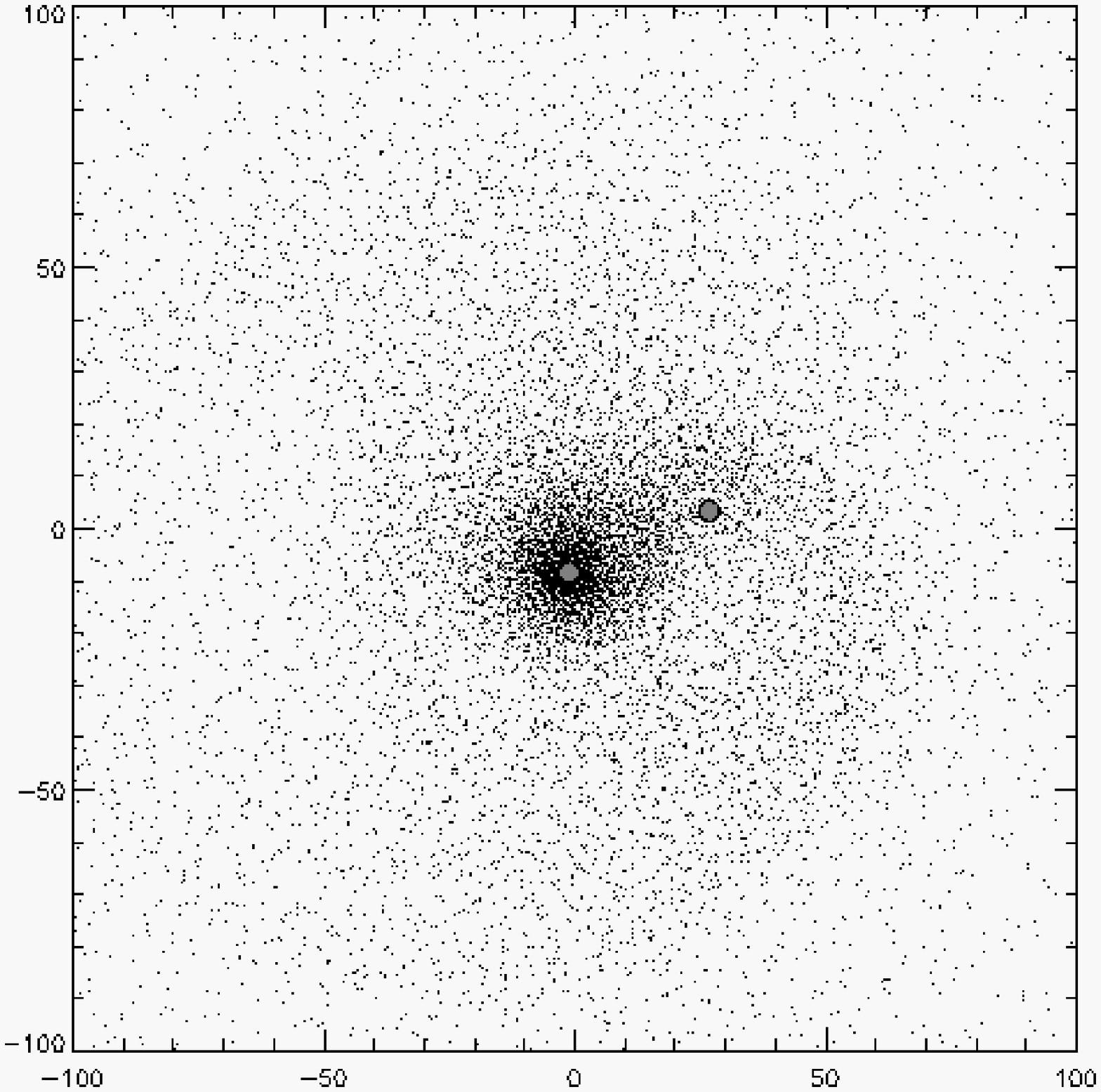}
\includegraphics[width=0.25\textwidth]{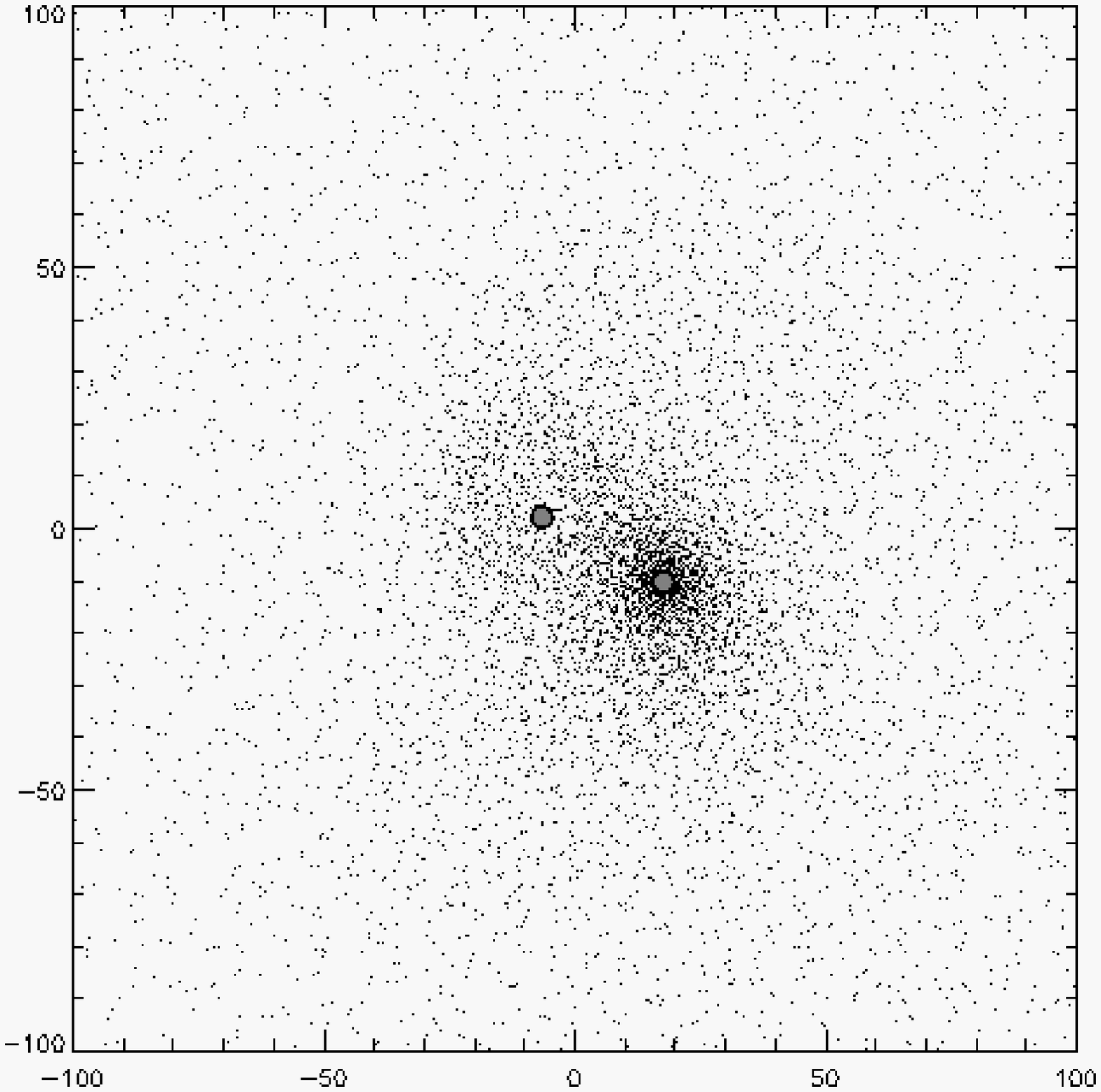}
\hfil
\caption{Time sequence of SPH particle plots for a common envelope between a $0.9M_{\sun}$ red giant and a $0.25M_{\sun}$ white dwarf. The small white dots show individual SPH prarticles, while the grey filled circles show the degerate core of the red giant (center) and the companion (right, on the surface). The individual plots are spaced 40 days apart from 0 to 200 days. Note how the companion transfers energy and angular momentum into the envelope, which is transported outward in a spiral density wave. At the end of this simulation, the binary has almost stalled its evolution, as there is only little material left at its radius to interact with, and most of the envelope corotates with the companion.
\label{f.WD025}}
\end{figure}

\section{SPH Simulations}

\begin{figure}
\hfil
\includegraphics[width=0.25\textwidth]{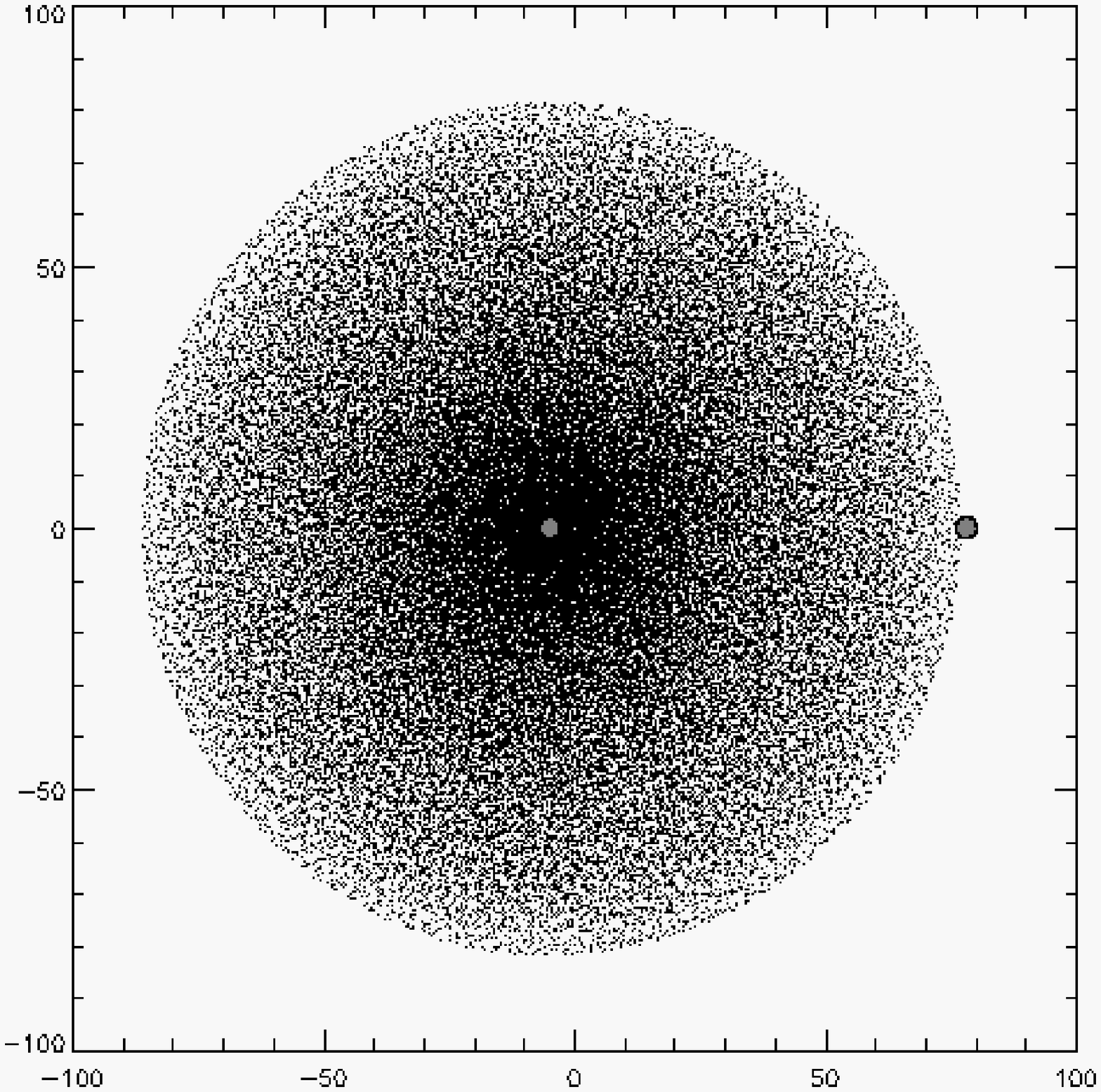}
\includegraphics[width=0.25\textwidth]{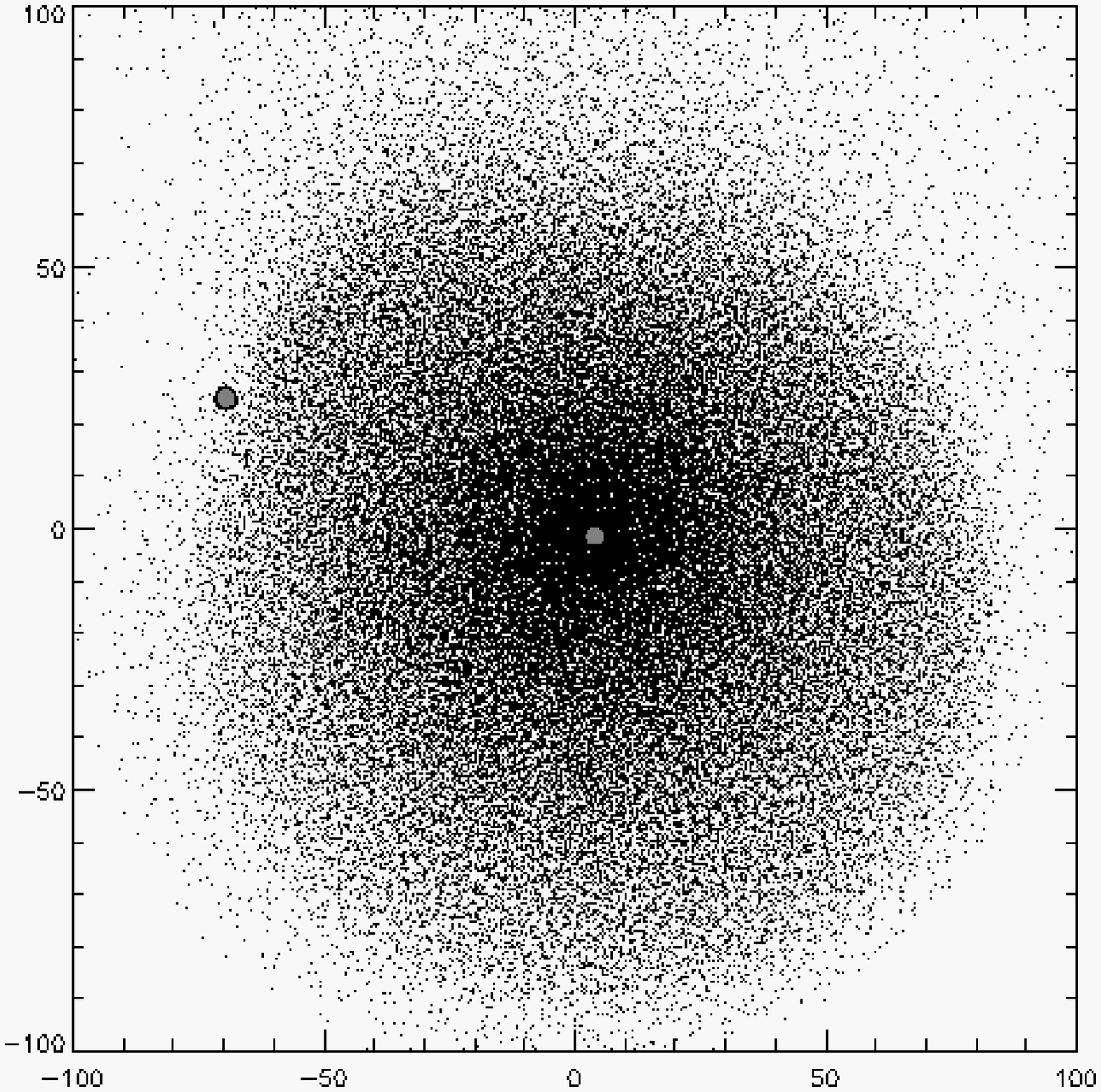}
\includegraphics[width=0.25\textwidth]{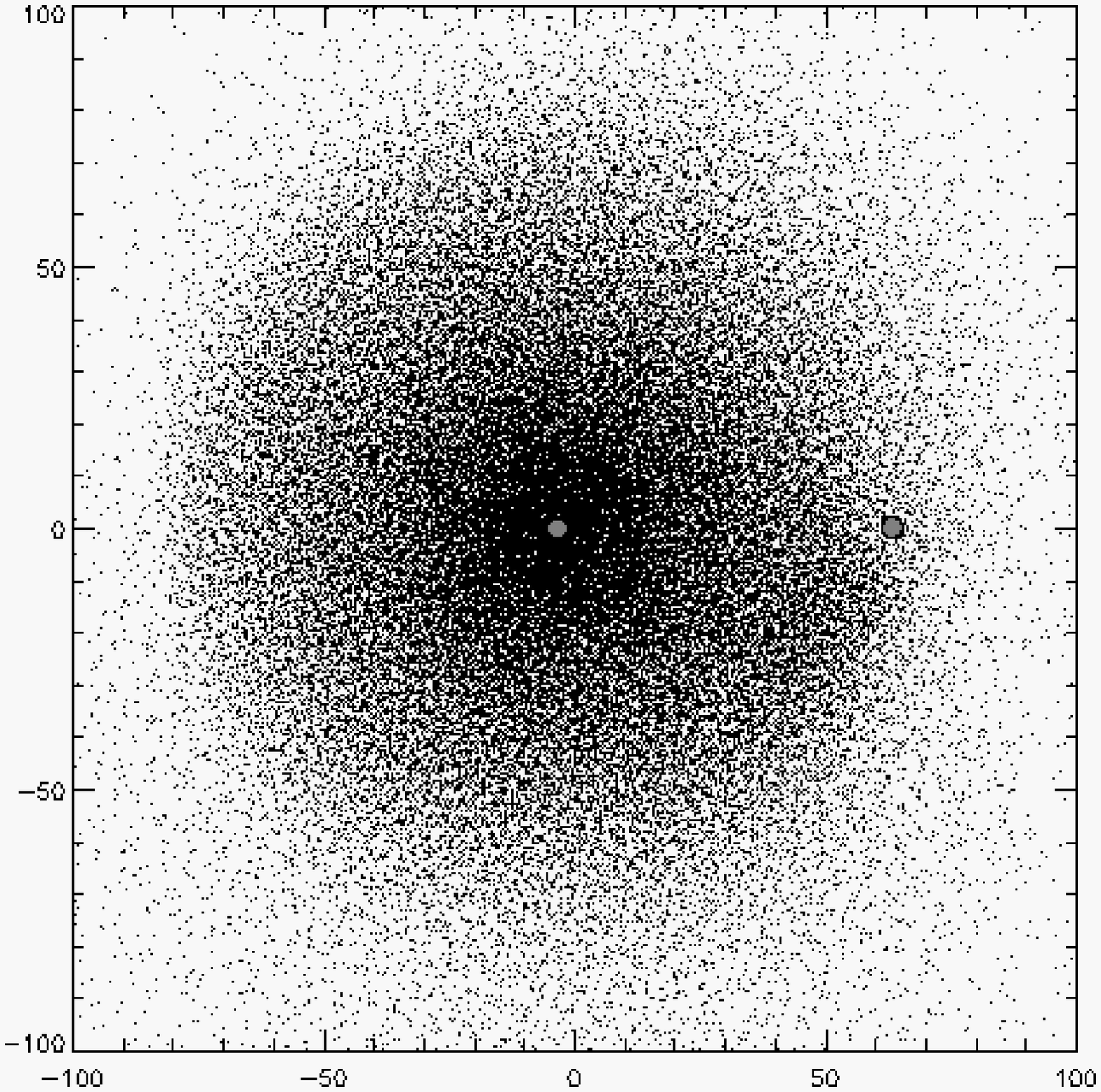} 
\hfil \\
\null\hfil 
\includegraphics[width=0.25\textwidth]{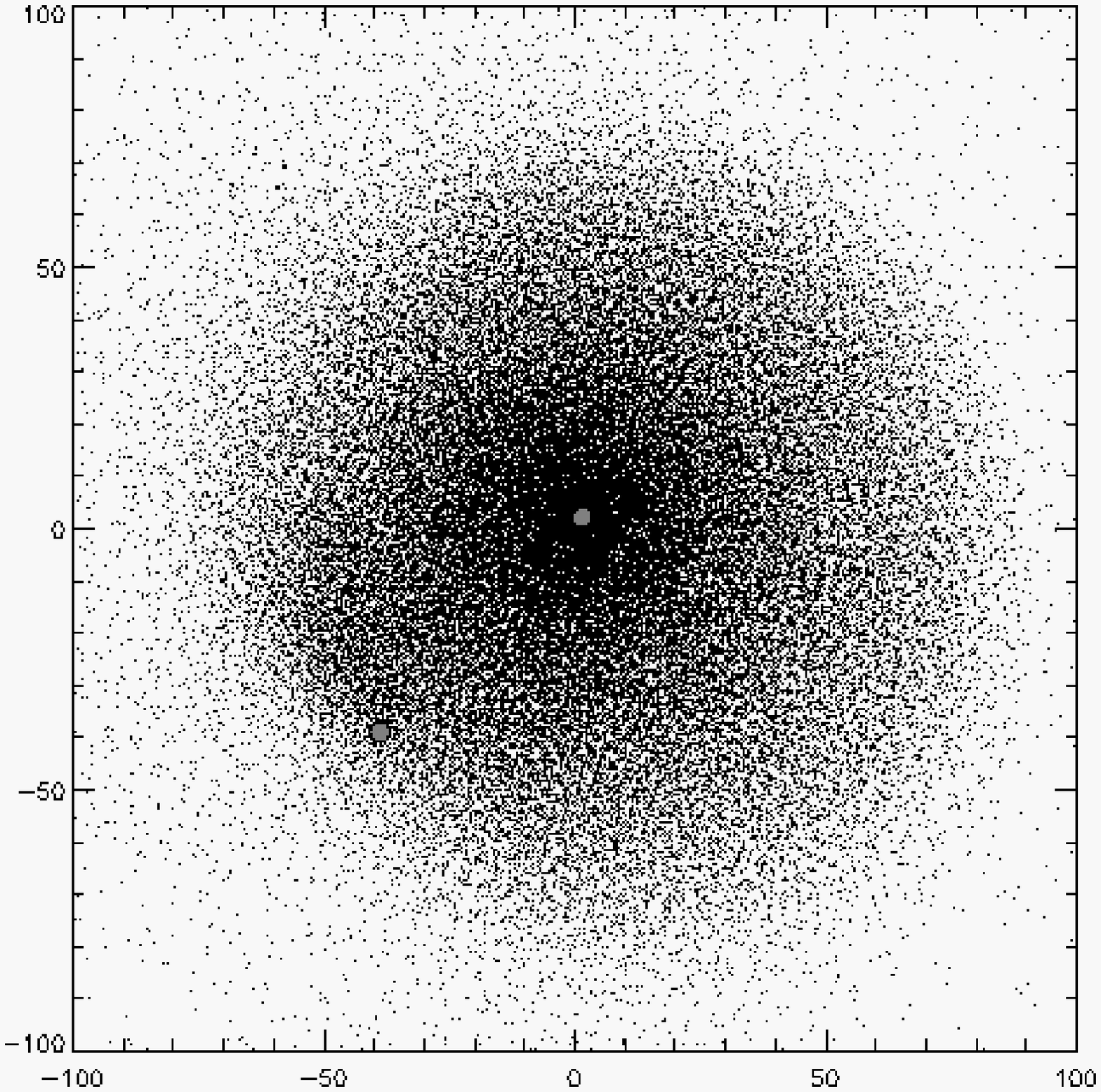}
\includegraphics[width=0.25\textwidth]{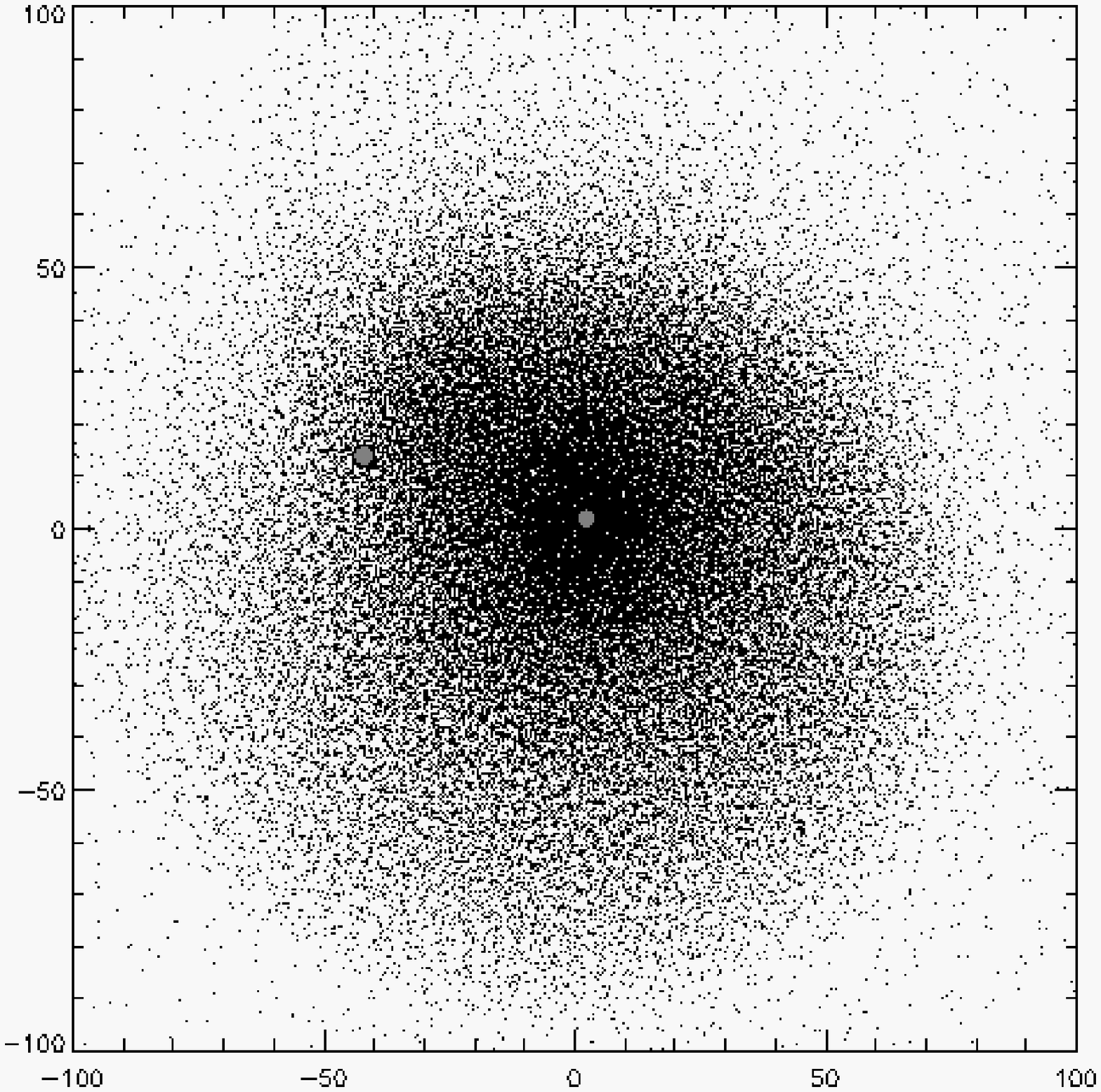}
\includegraphics[width=0.25\textwidth]{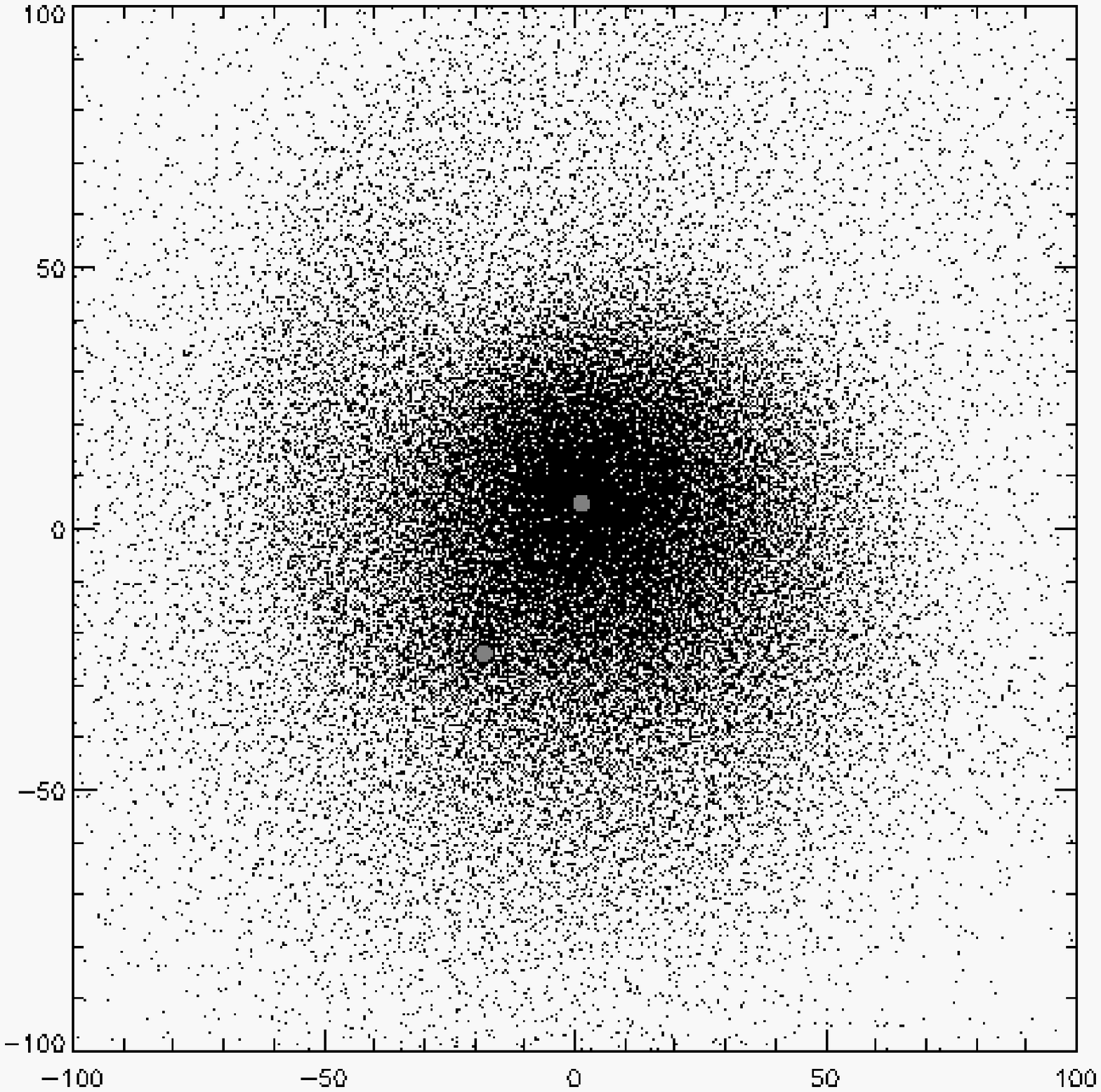}
\hfil \\
\null\hfil 
\includegraphics[width=0.25\textwidth]{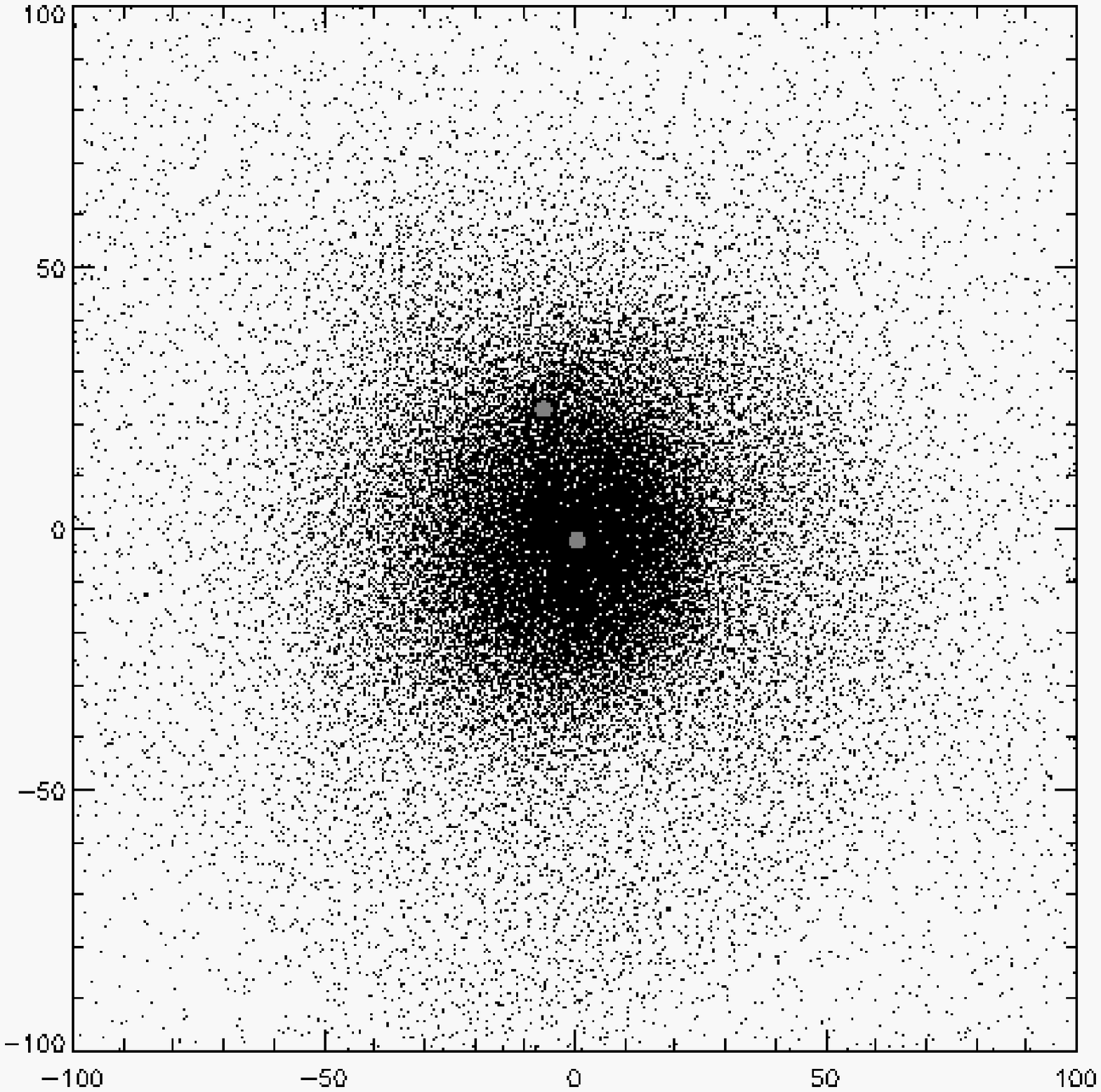}
\includegraphics[width=0.25\textwidth]{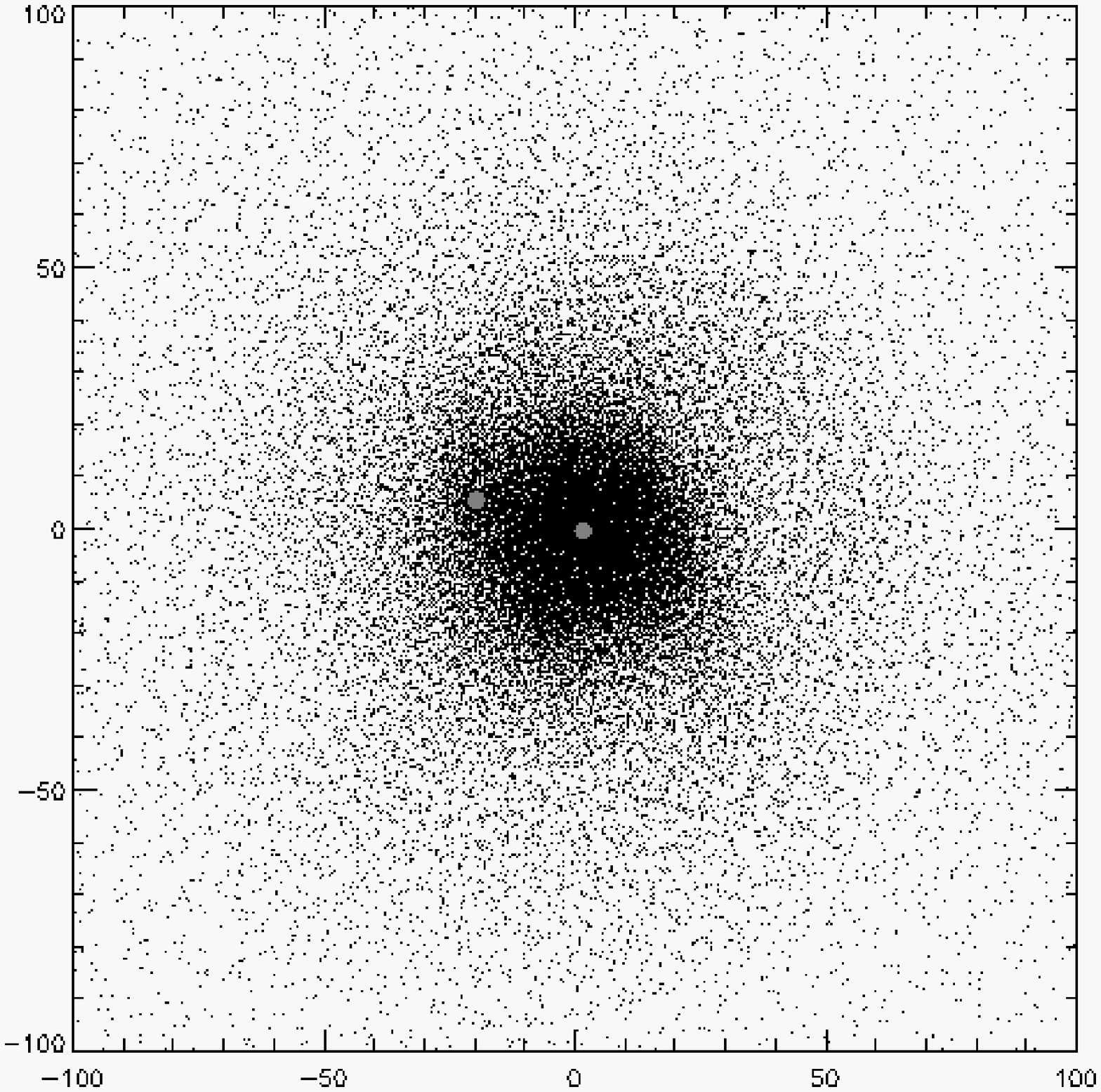}
\includegraphics[width=0.25\textwidth]{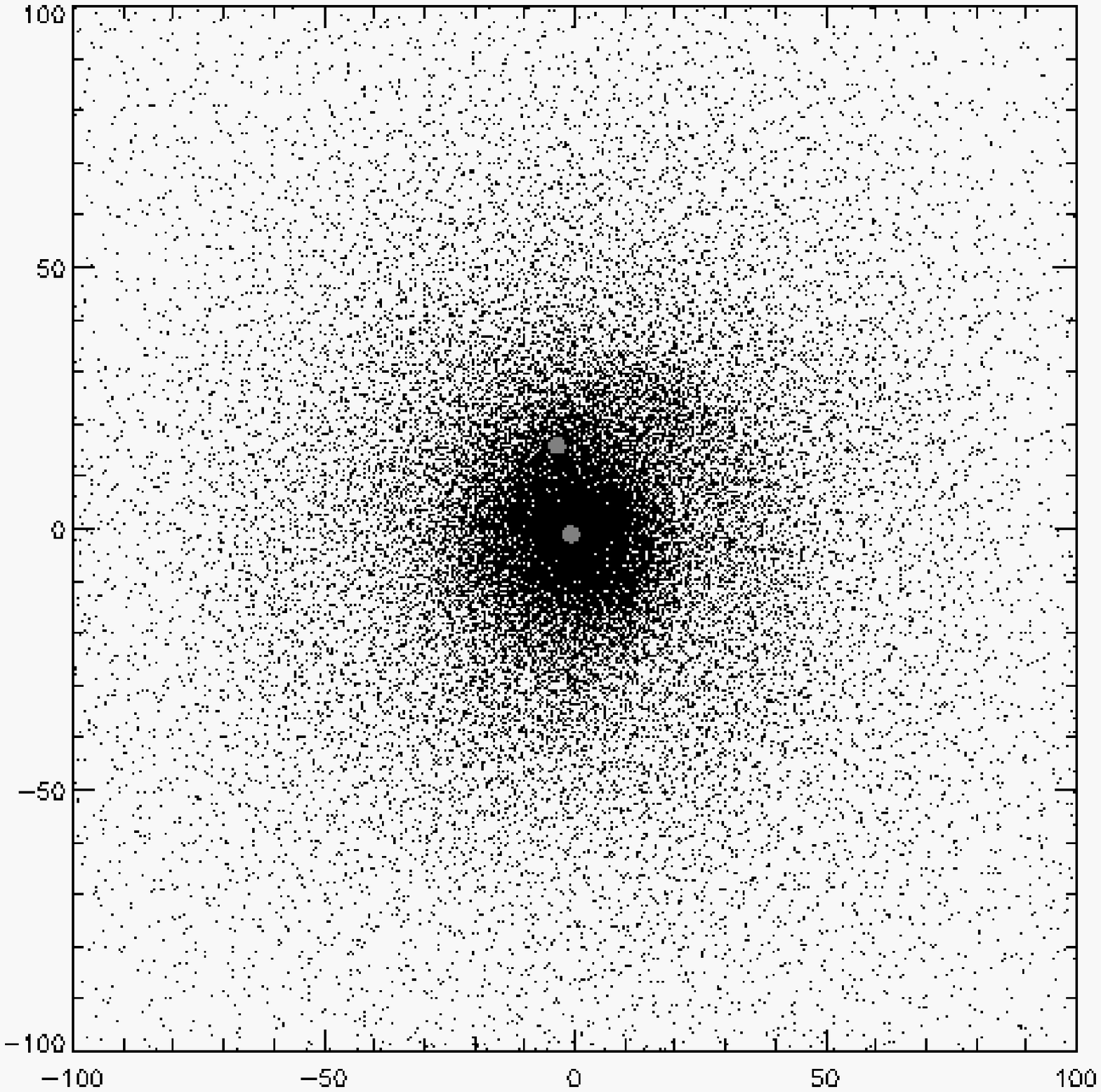}
\hfil
\caption{Same as Figure \ref{f.WD025}, but for a $0.05M_{\sun}$ brown dwarf companion. The images are also spaced 40 days apart, but extend until 320 days of evolution. Note how the interaction is much gentler, as the companion can transfer less energy and angular momentum into the envelope. The companion sinks in much deeper, and is in fact still sinking rapidly at the end of the evolution. The most energetic phase of this interaction is yet to come. \label{f.BD005}}
\end{figure}

We use the Smooth Particle Hydrodynamics (SPH) technique to simulate the CE evolution of a $0.9M_{\sun}$ RG interacting with two different companions: a medium mass companion ($0.25M_{\sun}$ WD, Figure \ref{f.WD025}) and a low mass companion ($0.05M_{\sun}$, Figure \ref{f.BD005}). Both simulations start with the companion in a circular orbit directly on the surface of the non-corotating star at $83R_{\sun}$. We use 100k SPH particles in these low-resolution test runs, still a factor of 2 higher than the last SPH simulations on this topic \citep{RasioCE}.

\begin{figure}
%\begin{center}
\hfil
\includegraphics[width=0.3\textwidth]{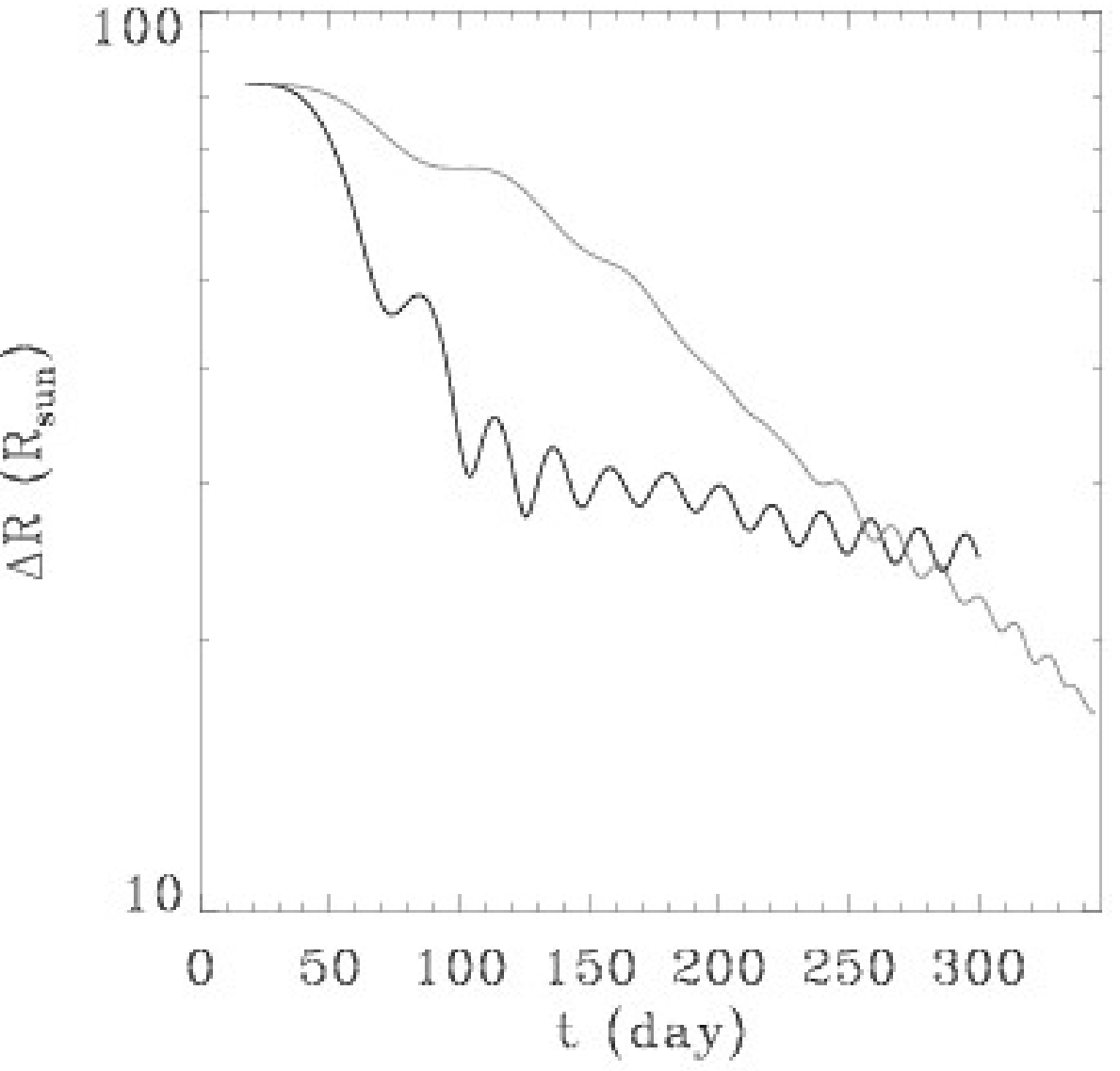}
\includegraphics[width=0.3\textwidth]{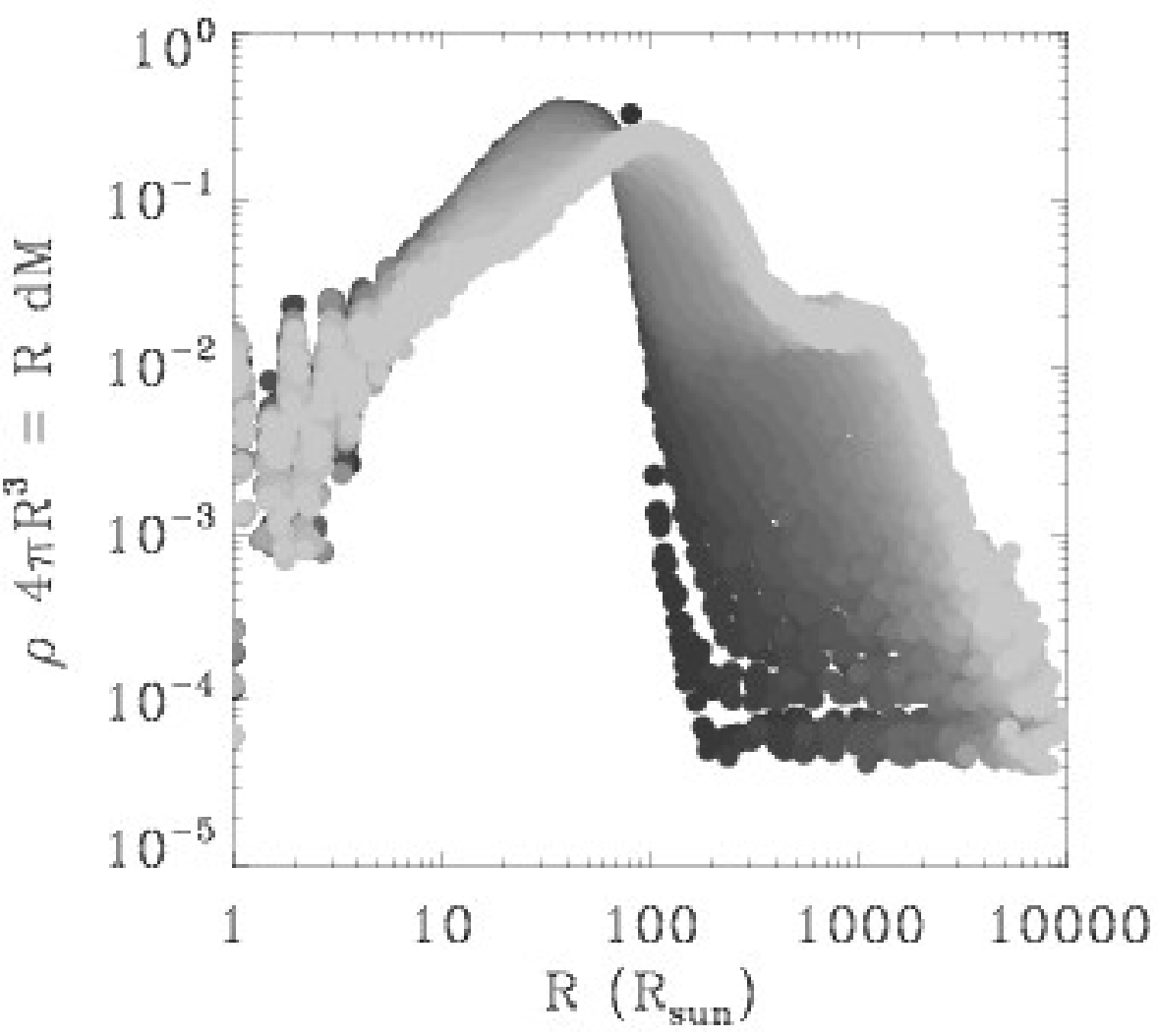}
\includegraphics[width=0.3\textwidth]{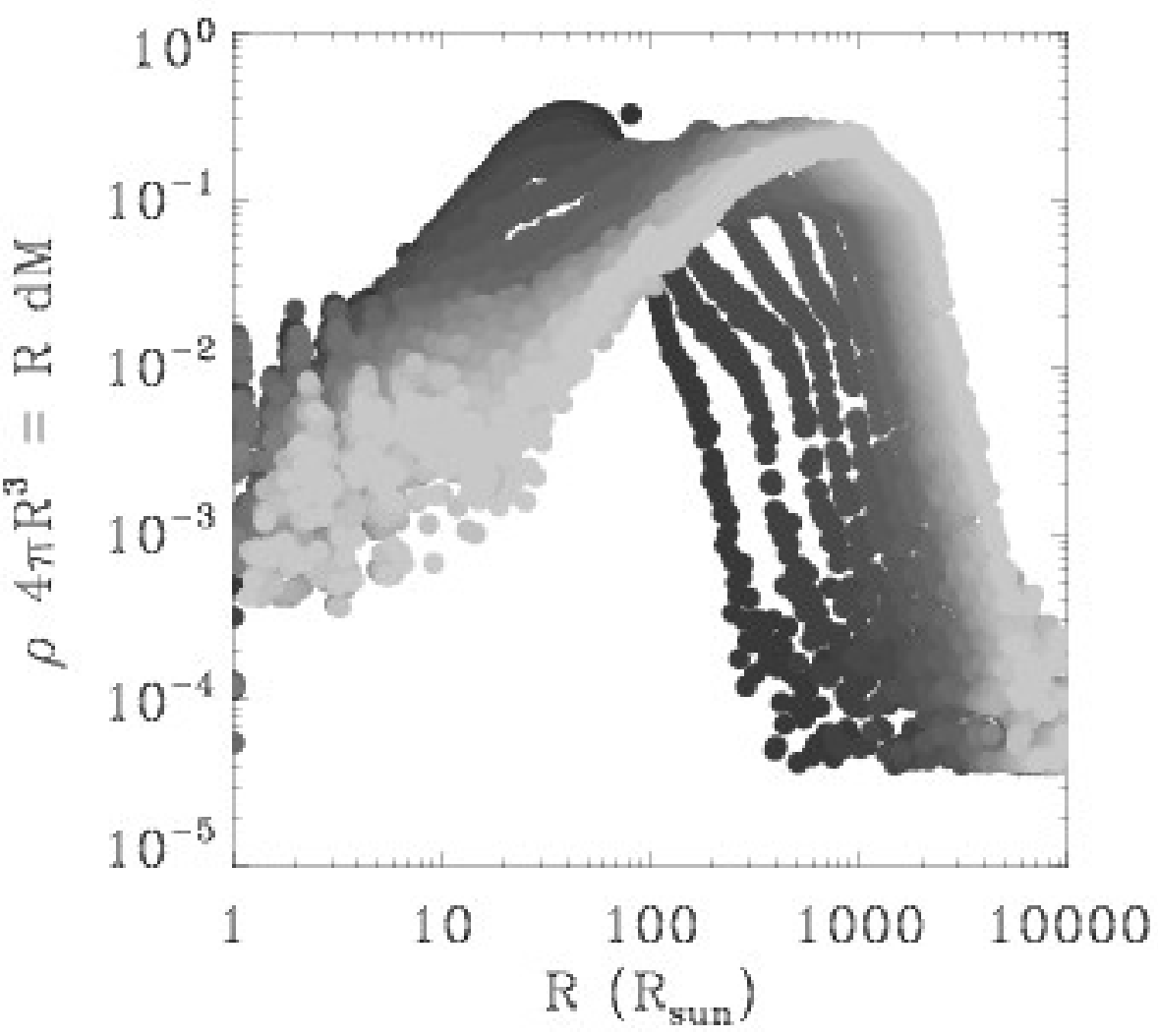}
\hfil
\caption{{\it Left}: Separation between RG core and companion as a function of time for the for the $0.05M_{\sun}$ (grey) and $0.25M_{\sun}$ companion (black). {\it Center and Right}: Mass distribution as a function of time for the $0.05M_{\sun}$ (left) and $0.25M_{\sun}$ companion (right). The area under the curve is proportional to the mass at that radius. Colors indicate the evolved time, ranging from 0 (darkest) to 300days (lightest). The higher-mass companion expels the envelope much faster with the bulk of the mass already being at around $1000R_{\sun}$. For the lower-mass companion on the other hand, most of the mass still hangs around $100R_{\sun}$. Material that is being ejected later at a higher speed has to ``plow'' through this wall first, potentially changing the dynamics of the evolution.\label{f.r_dm}}
%\end{center}
\end{figure}

Note how the higher mass WD companion is able to shed the outer layers of the star very quickly. Within only about 100 days, the binary orbit shrinks to a little more than a third ($30R_{\sun}$) of the initial radius (Figure \ref{f.r_dm}, left). However, after that, the evolution slows down significantly. Since part of the envelope has already been shed, there is less envelope material at this separation to transfer energy to. Additionally, the envelope slowly starts to corotate with the WD, making it even more difficult for the companion to interact. The lower-mass BD companion on the other hand interacts more gently with the envelope. Due to the lower orbital energy and angular momentum of the system, the companion cannot shed the outer layers as quickly. The BD slowly but steadily sinks deeper into the RG's envelope and is still sinking quickly at the current end of the simulation at around 350 days. We expect the BD to sink as deep as $0.65R_{\sun}$, as suggested by the recent observation of a similar BD-WD binary by \citet{MaxtedRGBD}, which we are trying to mimic here. 
Another major difference between the two simulations is the location of the ejected envelope material. The WD companion interacts violently with the outer layers of the RG and quickly moves the bulk of the material beyond $1000R_{\sun}$ (Figure \ref{f.r_dm}, right). The BD on the other hand ejects the material more gently, and the bulk still ``sits'' just beyond $100R_{\sun}$, just beyond the original surface of the RG (Figure \ref{f.r_dm}, middle), and there is some evidence that some of it may fall back down on the star again. Thus we expected the subsequent evolution of both cases to behave quite differently, as the most energetic phase of the BD interaction is still to come as it sinks deeper into the potential well. Any material that will be ejected from then on will have to first ``plow'' through the wall of cooled material that has accumulated around the star. 

In a forthcoming paper, we will explore the dynamics of CE evolution with higher resolution SPH simulations, putting particular emphasis on the neglected parameter space for low-mass companions. 

%\acknowledgements %%% Text of acknowledgements runs on after this command.

%This work was carried out in part under the auspices of the National
%Nuclear Security Administration of the Department of Energy (DOE) at Los Alamos
%National Laboratory and supported by contract DE-AC52-06NA25396. LA-UR-07-6063.

%%% THE BIBLIOGRAPHY
%%%
%%% CONSULT SECTION 3 OF "INSTRUCTIONS FOR AUTHORS" FOR HOW TO USE NATBIB.
%%% AUTHORS ARE ENCOURAGED TO USE EITHER THE "THEBIBLIOGRAPY" ENVIRONMENT
%%% BY UNCOMMENTING (DELETING THE "%" SYMBOL) THE COMMANDS BELOW, OR BY
%%% USING THE BIBTEX ENVIRONMENT. TO FIND OUT WHICH IS APPLICABLE TO YOUR
%%% CONTRIBUTION, CONSULT THE VOLUME EDITORS FOR YOUR PROCEEDINGS.
%%%

\bibliographystyle{apj}
\bibliography{/Users/diehl/papers/bibtex/allreferences.bib}

%\begin{thebibliography}{}
%\bibitem[]{}
%\bibitem[]{}
%\bibitem[]{}
%\bibitem[]{}
%\bibitem[]{}
%\bibitem[]{}
%\bibitem[]{}
%\bibitem[]{}
%\bibitem[]{}
%\bibitem[]{}
%\bibitem[]{}
%\bibitem[]{}
%\end{thebibliography}

\end{document}